\documentclass[aps,twocolumn,showpacs]{revtex4}
\usepackage{graphicx}
\usepackage{color}
\usepackage{amssymb}
\usepackage{amsmath}

\begin{document}

\title{Dynamics of a charged particle in a dissipative Fermi-Ulam model}

\author{Denis Gouv\^ea Ladeira$^1$ and Edson D.\ Leonel$^2$}
\affiliation{$^1$ Departamento de F\'isica e Matem\'atica - 
UFSJ - Univ Federal de S\~ao Jo\~ao del-Rei, \\
Rod. MG 443, Km 7 - Fazenda do Cadete -
36420-000 - Ouro Branco - MG - Brazil\\
$^2$ Departamento de F\'isica - UNESP - Univ Estadual Paulista,\\
Av.24A, 1515 - Bela Vista - 13506-900 - Rio Claro - SP - Brazil}

\date{\today}\widetext


\begin{abstract}

The dynamics of a metallic particle confined between charged walls is
studied. One wall is fixed and the other moves smoothly and periodically in
time. Dissipation is considered by assuming a friction produced by the contact
between the particle and a rough surface. We investigate the phase space of
the simplified and complete versions of the model. Our results include (i) 
coexistence of islands of regular motion with an attractor located at the low
energy portion of phase space in the complete model; and (ii) coexistence of
attractors with trajectories that present unlimited energy growth in the
simplified model.


\end{abstract}

\maketitle

\section{Introduction}
\label{introduction}

Processes of accelerating particles have been a subject of large interest in
recent years. Billiards are systems widely used to study these processes due
to their adaptability to several situations by exploiting different
mechanisms. A billiard is a system where a particle (or more) collides with
boundaries while it interacts, or not, with some potentials. These boundaries
can be static or time dependent, as well as the potentials. Billiards systems
usually present, mathematically speaking, a nonlinear term and their phase
spaces exhibit a rich variety of behaviours including, but not limited to,
chaotic trajectories and quasi-periodic orbits. The studies include: (i) the
non-dissipative case, where dissipation due to friction, drag or inelastic
collisions is absent, and; (ii) the dissipative case, where {\it usually} the
islands of regular motion give place to attractors and, depending on
parameters, the chaotic sea becomes a chaotic attractor or eventually, a
chaotic transient. Moreover, depending on perturbation, the particle can
present a power-law energy growth, that is suppressed by weak linear
dissipation, or a robust exponential energy growth. 

The phenomena where a classical particle acquires unbounded energy when it is
excited by an external driving structure is called Fermi acceleration (FA).
The studies of FA are risen since the first half of 20th century, when high
energy cosmic rays were observed. The basic mechanism of acceleration was
proposed by Enrico Fermi \cite{Fermi} and it consists of the interaction of
the cosmic ray with time dependent magnetic fields generated by the activity
of galaxies nuclei. Such interaction was claimed to be the responsible for
giving the high energy to the particle.

Fermi's idea can be placed in terms of a classical model. Indeed it is similar
to the problem of a particle colliding against a rigid and a moving walls. The
particle corresponds to the cosmic ray while the moving wall represents the
driving provided by the time dependent magnetic fields. Ulam \cite{ulam} was
who proposed this model, where a particle moves in absence of any external
field and dissipation between collisions with the walls. The fixed wall works
as a mechanism to re-inject the particle back for further collisions with the
moving wall. Such a model, known as Fermi-Ulam model (FUM) is described by a
two-dimensional and area preserving mapping. The amplitude of oscillation of
the moving wall defines the strength of the nonlinear term. For non null
amplitude of oscillation, the phase space of the FUM presents mixed structure,
where regions of chaotic motion coexist with Kolmogorov-Arnold-Moser (KAM)
islands and invariant tori, also called as invariant spanning curves or
invariant rotational curves. Moreover, these spanning curves limit the energy
gain for a chaotic orbit, prohibiting an unlimited diffusion in the velocity
axis. Therefore, FA is not achieved in FUM, as it was  discussed by 
Lichtenberg {\it et al.} \cite{10.1016:0167-2789(80)90027-5.pdf}. 

A variation of FUM called as bouncer regards the dynamics of a classical
particle moving in the presence of a constant gravitational field and
suffering elastic collisions with a periodically moving wall. The re-injection
mechanism is now produced by the gravitational field. For specific sets of
control parameter and initial conditions, the bouncer model
\cite{BouncerOrig_Pustyl,bouncer01} indeed presents FA. For small non null
amplitude of oscillation of the platform the phase space presents small
regions of chaotic motion, islands of regular motion and invariant spanning
curves that prevent the particle to gain unlimited energy. Consistently with
KAM theorem, invariant spanning curves are continually destroyed as the
amplitude of oscillation increases until all of them disappear. As a
consequence, the chaotic component of phase space becomes open and the average
velocity of the particle grows unlimitedly leading to FA. Lichtenberg {\it et
al.} \cite{10.1016:0167-2789(80)90027-5.pdf,LichtenbergLiebermanBook}
demonstrated that the bouncer model is globally equivalent to the Chirikov's
standard map, explained the origin of FA and, moreover, obtained the
combinations of initial conditions and values of parameter where the
accelerating modes are observed in the bouncer model. 

Concerning FA in two-dimensional time dependent billiards, it was conjectured
by Loskutov, Ryabov and Akinshin that a sufficient condition for the
observation of FA is the presence of chaotic motion of the particle in the
static boundaries version of such a billiard (existence of positive Lyapunov
exponents on a nontrivial region of phase space) \cite{LRA.01,LRA_Conjctr}. 
This conjecture, known as LRA conjecture, was confirmed in different billiards
\cite{LRA_conf.01,ref8,ref9,ref10}. Recently it has been reported
FA in an harmonically driven version of the integrable static elliptical
billiard \cite{LDS=LRA+}. The invariant manifolds observed for the static
version become penetrable in the driven model, generating a laminar dynamics.
Moreover, the driving replaces the separatrix by a chaotic layer where the
motion is stochastic, introducing a diffusion process to the velocity and
leading the particle to present FA. After that, the LRA conjecture was
amended \cite{10.1103:PhysRevLett.104.224101.pdf}, where the existence of a
chaotic component was replaced by the existence of a heteroclinic orbit in
phase space. 

Recently, it has been studied the aspects of a nonstandard
exponential energy growth
\cite{10.1103:PhysRevE.81.056205.pdf,10.1103:PhysRevLett.106.074101.pdf}. This
phenomena was numerically observed in a slitted rectangular billiard with a
bar oscillating smoothly, where it was demonstrated that the origin of the
exponential FA is due to the transition of the particle between the invariant
spanning curves observed in phase space of the FUM
\cite{10.1103:PhysRevE.81.056205.pdf}. Also, the aspects of a robust
exponential FA were obtained analytically and numerically in a class of
systems that consists of billiards where a particle collides against 
time dependent boundaries that are deformed, separated and then reconnected 
\cite{10.1103:PhysRevLett.106.074101.pdf}. Moreover, it has been
mathematically demonstrated the existence of orbits with energy growth rate
ranging from power law to exponential FA in non-autonomous billiards
\cite{10.1007:s00220-008-0518-1.pdf,10.1088:1751-8113:41:21:212003.pdf,
10.1063:1.4736542.pdf}. 

Other important observation on the state of the art of the problem is that FA
has demonstrated to be not a robust phenomena under the presence of inelastic
collisions \cite{LRAd.02,innel.01} and under viscous drag force
\cite{drag.01}. However few studies are known about consequences of friction
due to the slip of a body on a surface on FA process. When friction acts on a
particle in a FUM where a wall moves according to saw-tooth function, FA can
either persist or not depending on initial conditions and parameters
\cite{fric.01}. 

Before we start to talk about this work, it is interesting to say that 
the study of the billiards class is of interest specially because the versatility of practical description 
to a wide variety of systems, such as 
waveguide and rippled channel \cite{waveguide,RippledChannelWaveguide}, 
magnetic field lines in toroidal plasma devices \cite{MagneticFieldLines_ToroidalPlasmaDev}, 
cold atoms \cite{ColdAtoms}, channel flows \cite{ChannelFlows.01,ChannelFlows.02}, escape processes 
\cite{escape.01,escape.02}  and transport properties \cite{Transport.01,Transport.02}. 

In this work we study a modified version of Fermi-Ulam model where a charged 
particle moves under dissipation due to friction and it is accelerated by an 
electric field between metallic walls. We are seeking to understand and
describe some dynamical properties of the system either as function of the
time as well as according to the control parameters. We regard that the
oscillating wall moves smoothly and we study both the simplified and complete
models. Under specific situations the model under study is a prototype for FA
and in the scenario of null electric field the model corresponds to a FUM
with friction. Depending on parameters, the simplified model presents a
coexistence of attractors with trajectories that present FA. Moreover, in the
complete model we observe numerically the coexistence of KAM islands with
trajectories that evolve to null velocity, even for null electric field. 

The paper is organized as follow. In the next section we present the model 
and we discuss the results of both simplified and complete versions. 
In section \ref{conclusions} we provide the conclusions and 
final remarks. 

\section{The model}
\label{model}

It is well known that charged particles are accelerated by electric fields. In
this paper we propose a modification of the FUM where the walls are assumed to
be metallic and sufficiently large and massive. They are charged with opposite
charges generating an electric field $\vec{E}$ that acts on a charged particle
between them. For instance, let us say that $\vec{E}$ is in the direction
$x$, $\vec{E}=E\hat{i}$. At each collision with the left wall the particle
acquires charge $+q$, $q\ge 0$, and an electric force $F_e=qE$ acts on the
particle forcing it to move to the right. Similarly, the particle acquires
charge $-q$ when it collides against the right wall and it becomes under
action of an electric force $F_e=-qE$. The left wall oscillates smoothly with
amplitude $\epsilon$ and frequency $\omega$ according to the expression
$x=\epsilon\cos(\omega t^\prime+\phi_0)$, where $t^\prime$ is time and
$\phi_0$ is an initial phase. The right wall is fixed at position $x=l$. We
regard the situation where collisions are elastic, so the absolute value of
the velocity of the particle relative to the walls is not affected by
collisions. 

Additionally, we consider that the particle moves under action of a friction
force $F_f=\pm\sigma N$, due to the friction of it on a rough surface. The
quantity $\sigma\ge 0$ corresponds to the kinetic coefficient friction and
$N\ge 0$ is the magnitude of a normal force. Because $F_f$ opposites the
motion (velocity) of the particle, $F_f\ge 0$ when the particle moves to the
left (velocity $v<0$) and vice-versa. 

It is also important to define $\sigma_s>\sigma$ as the static friction
coefficient. So $|F_s|=\sigma_s N$ is the maximum absolute value of the static
friction force that acts on the particle when it is stopped. The signal of
$F_s$ depends on tendency of motion. So if the particle stops after colliding
the right wall, then $F_s\ge 0$. Similar situation occurs when the particle
stops after colliding the moving wall. 

Therefore, when the particle is moving between the walls, it is under action
of a force $F=F_e+F_f$. When the particle stops it is under a force
$F=F_e+F_s$. So, depending on signal and strength of electric and friction
forces the particle can gain or loose energy while it moves between two
collisions. In terms of electric, static and dynamical friction forces we
define the following set of parameters: $A=(qE-\sigma N)/(\omega^2 lm)$, 
$B=(qE+\sigma N)/(\omega^2 lm)$, $C=(qE-\sigma_s N)/(\omega^2 lm)$, where $m$
is the mass of the particle. Moreover, it is appropriate to define the
following set of dimensionless variables: $X=x/l$, $V=v/(\omega l)$, $t=\omega
t^\prime$ and $\phi=\omega t^\prime+\phi_0$. We define the dimensionless
amplitude of oscillation by the expression $\varepsilon=\epsilon/l$. 

From the above definitions, we have that the parameter $A$ represents the
competition between the electric and dynamic friction forces. If $A>0$, then
the electric force is greater than the dynamical friction force (and
vice-versa). Similarly, the parameter $C$ represents the competition between
the electric and static friction forces. Therefore if the particle reaches the
rest and $C<0$, then the maximum static friction force is greater than the
electric force. If this situation occurs in the region where the moving wall
oscillates, then the particle remains in rest and waiting a collision with the
moving wall. However if the particle stops somewhere outside the region of
oscillation of the moving wall, then its dynamics is over. If $C>0$, then the
particle: (i) does not stop after colliding the fixed wall or; (ii) it stops
instantaneously if its velocity is negative after a collision with the moving
wall.

It is interesting to observe that the situations where $A<0$ and $C<0$ can be
interpreted as null electric force, $qE=0$, and the model corresponds to a FUM
with dissipation due to friction. The non null electric field is important to
produce non-negative values of $A$ and $C$. 

Let us discuss initially the dynamics of the particle regarding a
simplification of the model and, after that, we present the results of the
complete model. 

\subsection{The simplified map}

In the simplified version of the model the oscillating wall imparts momentum
to the particle. However, it is assumed that it takes a stationary position.
So the time interval spent by the particle in the travel between collisions 
does not depend on the phase of the oscillating wall. The simplified versions
are widely studied because they are useful under some aspects. For example,
they allow to obtain analytical results that can be extended to the complete
versions and, usually, they speed up greatly the simulations when compared to
the corresponding complete versions.

If at an instant $t_n$ the particle hits the moving wall and acquires velocity
$V_n$, then the dynamics of the system after subsequent collisions is obtained
by the following two-dimensional map 
\begin{eqnarray}
\label{SimplifiedMap}
T:\left\{
\begin{array}{l}
V_{n+1}=|-\sqrt{V_n^2+4A}+2\varepsilon\sin\phi_{n+1}|
\\
\phi_{n+1}=(\phi_n+\Delta t_{n+1})\mod 2\pi
\end{array}
\right.
\end{eqnarray}
where $\phi_n=t_n +\phi_0$ is the phase of the wall at instant $t_n$,
and 
\begin{eqnarray}
\label{Deltat}
\Delta t_{n+1}=\frac{1}{A}\left[\sqrt{V_n^2+4A}-V_n\right]
\end{eqnarray}
is the interval of time between two collisions with the wall. The absolute
value bars are required to avoid the particle to leave the region between
the walls. If the inequality $V_n^2+4A<0$ is satisfied, then the particle
stops between the walls and the the simulation for this trajectory is
terminated.

Let us now proceed with the investigation of the fixed points and their
stability, which will be important latter on along the paper. The fixed
points are obtained from map (\ref{SimplifiedMap}) by applying the 
conditions $V_{n+1}=V_n=V^*$ and $\phi_{n+1}=\phi_n=\phi^*$. So we
obtain the coordinates $(\phi^*,V^*)$ of the fixed points in phase space as
\begin{eqnarray}
\label{fixedpoints}
\begin{array}{l}
V^*=\frac{1}{\pi i}-\pi i A\\
\phi^*=\arcsin\left(\frac{\pi i A}{\varepsilon}\right),
\end{array}
\end{eqnarray}
where $i$ is a non null positive integer. For each value of $i$ and
$\varepsilon$ the above expression furnishes, in the interval
$-\varepsilon/(\pi i)<A<\varepsilon/(\pi i)$, a pair of fixed points with the
same velocity and different values of phase. Let us nominate $\phi^*_1$ and
$\phi^*_2$ the two possible values of $\phi^*$ associated to each value of
$V^*$ given by Eq. (\ref{fixedpoints}). If the absolute value of $A$
increases, or $\varepsilon$ decreases, then the maximum value of $i$ decreases
and, therefore, it decreases the maximum number of fixed points. So, for
$|A|/\varepsilon>\pi^{-1}$ the phase space does not present any fixed point. 

To classify the fixed points according to their stability, we must obtain
the eigenvalues of the Jacobian matrix 
\begin{eqnarray}
\nonumber
J=
\left(
\begin{array}{ll}
\frac{\partial V_{n+1}}{\partial V_n} & \frac{\partial V_{n+1}}{\partial \phi_n}
\\
\frac{\partial \phi_{n+1}}{\partial V_n} & \frac{\partial \phi_{n+1}}{\partial \phi_n}
\end{array}
\right)
\end{eqnarray}
where 
\begin{eqnarray}
\begin{array}{ll}
\frac{\partial V_{n+1}}{\partial V_n}&=\frac{V_n}{\sqrt{V_n^2+4A}}-2\varepsilon\cos\phi_{n+1}\frac{\partial\phi_{n+1}}{\partial V_n},
\\
\frac{\partial V_{n+1}}{\partial \phi_n}&=-2\varepsilon\cos\phi_{n+1},
\\
\frac{\partial\phi_{n+1}}{\partial V_n}&=\frac{1}{A}\left(\frac{V_n}{\sqrt{V_n^2+4A}}-1\right)
\\
\frac{\partial\phi_{n+1}}{\partial\phi_n}&=1.
\end{array}
\end{eqnarray}
The eigenvalues $\Lambda$ are obtained from equation $\det(J-\Lambda I)=0$.
The procedure furnishes, for each fixed point, a pair of eigenvalues
$\Lambda_{1,2}=\frac{1}{2}\left({\rm Tr}J\pm\sqrt{({\rm Tr}J)^2-4\det
J}\right)$, 
where 
${\rm Tr}J=\frac{V_n}{\sqrt{V_n^2+4A}}-2\varepsilon\cos\phi_{n+1}\frac{\partial \phi_{n+1}}{\partial V_n}+1$
and 
\begin{eqnarray}
\det J=\frac{V_n}{\sqrt{V_n^2+4A}}.
\end{eqnarray}

\begin{figure}[htb]
\centerline{
\includegraphics[angle=0, width=7.5cm, height=4cm]{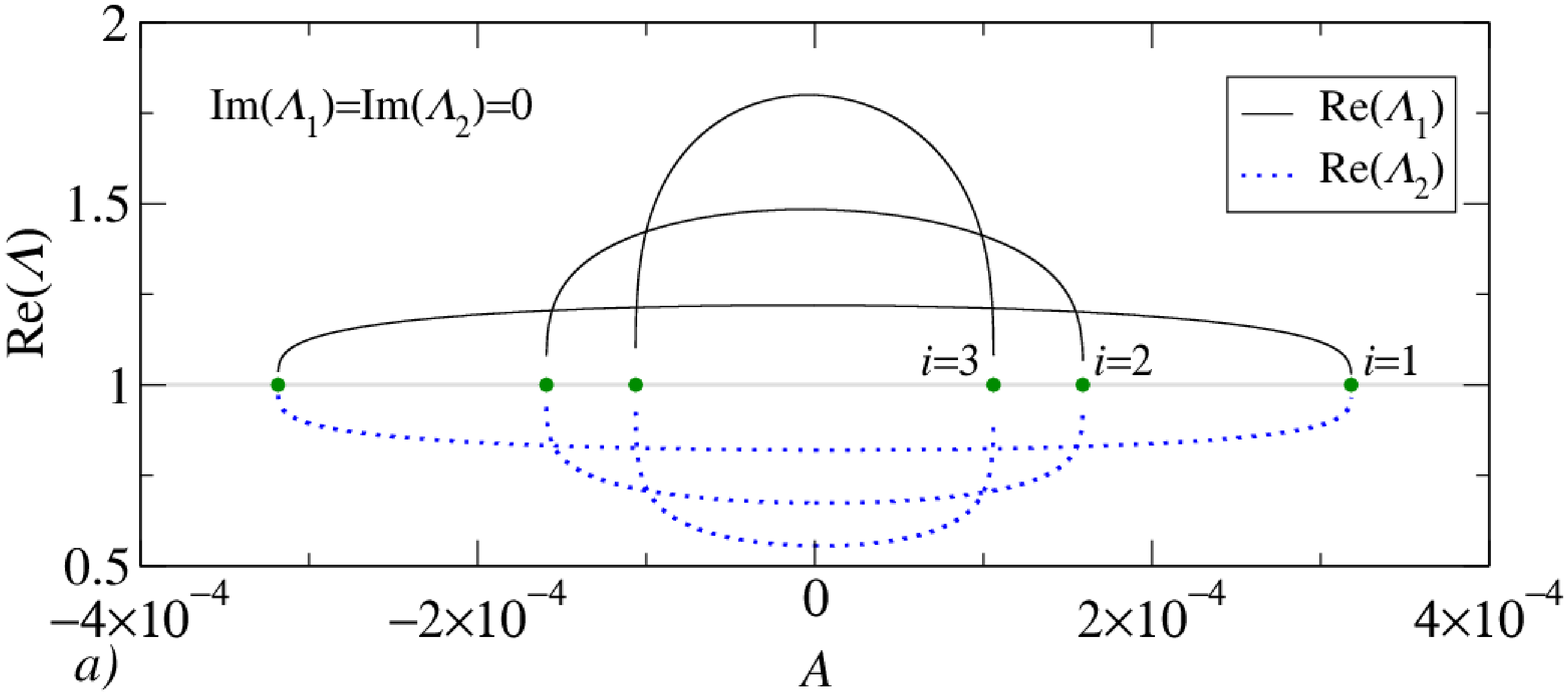}
}
\caption{Eigenvalues $\Lambda_1$ and $\Lambda_2$ associated to the fixed
points $(\phi^*_1,v^*)$.}
\label{eigenvaluesphi1}
\end{figure}

We evaluated the eigenvalues of $J$ regarding the fixed points given by Eq.
(\ref{fixedpoints}) as a function of $A$ for $\varepsilon=10^{-3}$. Figure
\ref{eigenvaluesphi1} illustrates the eigenvalues for the fixed points
$(\phi^*_1,V^*)$. The smaller the $|A|$ the greater the number of fixed
points. To simplify the visualization we include the eigenvalues of fixed
points with $i=1,2,3$. But decreasing the absolute value of $A$ we observe
similar curves for other values of $i$. Both $\Lambda_1$ and $\Lambda_2$ are
real quantities, and because $\Lambda_1\ge 1$ and $\Lambda_2\le 1$, the fixed
points $(\phi^*_1,V^*)$ are classified as saddle. 

\begin{figure}[htb]
\centerline{
\includegraphics[angle=0, width=7.5cm, height=5cm]{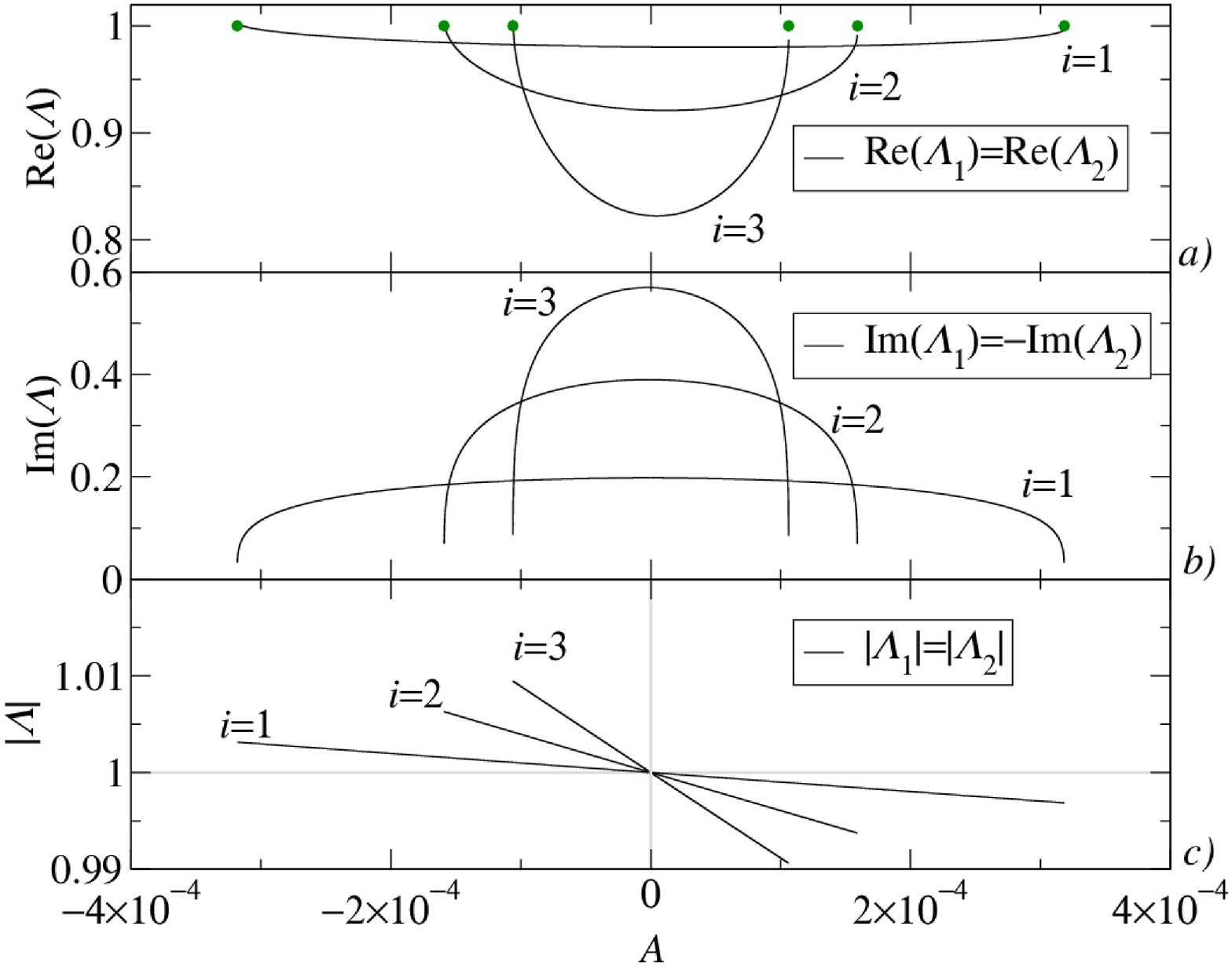}
}
\caption{Eigenvalues $\Lambda_1$ and $\Lambda_2$ associated to the fixed
points $(\phi^*_2,v^*)$: a) real part of $\Lambda_1$ and $\Lambda_2$, b)
imaginary part of $\Lambda_1$ ($=-{\rm Im}(\Lambda_2)$) and c) absolute
value of $\Lambda_1$ ($=|\Lambda_2|$).}
\label{eigenvaluesphi2}
\end{figure}

The fixed points $(\phi^*_2,V^*)$ have complex eigenvalues. Therefore these
fixed points are classified as spiral focus (attracting or repelling, as we
shall see). Fig. \ref{eigenvaluesphi2}a) illustrates the real part of
$\Lambda_{1,2}$, while Fig. \ref{eigenvaluesphi2}b) illustrates the imaginary
part of $\Lambda_1$ (${\rm Im}(\Lambda_1)=-{\rm Im}(\Lambda_2)$). The
absolute value of both $\Lambda_{1,2}$ is displayed at Fig.
\ref{eigenvaluesphi2}c). We observe that these fixed points are repelling
focus for $A<0$ ($|\Lambda_{1,2}|>1$) and they are attracting focus for $A>0$
($|\Lambda_{1,2}|<1$). 

From Figs. \ref{eigenvaluesphi1} and \ref{eigenvaluesphi2} and the above 
discussion we conclude that the phase space presents saddle-node 
bifurcations at $A=\pm\varepsilon/(\pi i)$. The bullets in such figures 
indicate the values of $A$ where some of these bifurcations occur. 

Let us now discuss the dynamical aspects of trajectories in phase space and
how these trajectories are organized in terms of the manifolds of the fixed
points of the system under study. Each saddle fixed point presents four
manifolds. Two of them are stable (attractive), in sense that trajectories on
them converge asymptotically to the saddle point, and two are unstable
(repulsive), where ICs generate trajectories that diverge from the saddle
point. Each manifold is constructed using the direction of the eigenvectors
$\eta$ at the saddle point, obtained from the expression $J\eta=\Lambda\eta$. 

The unstable manifolds are constructed by iterating a set of ICs defined
along a line with orientation of the corresponding eigenvector, near the saddle
point. The construction of the stable manifolds is slightly more
complicated because it requires the construction of the inverse of the map, which
furnishes the values of velocity, $V_n$, and phase of the wall $\phi_n$ from
the next collision, when velocity and phase are $V_{n+1}$ and $\phi_{n+1}$, 
\begin{eqnarray}
\label{mapainversosimplificado}
T^{-1}:\left\{
\begin{array}{l}
V_n=\sqrt{(V_{n+1}+2\varepsilon\sin\phi_{n+1})^2-4A}
\\
\phi_n=[\phi_{n+1}-\Delta t_{n+1}]~~{\rm~mod}~2\pi
\end{array}
\right.,
\end{eqnarray}
where $\Delta t_{n+1}$ is given by Eq. (\ref{Deltat}). Near the saddle point,
we define a set of points orientated with the stable eigenvectors as the 
ICs of the inverse map and obtain the stable manifolds. 

Let us discuss at first the situation where $A<0$. For $\varepsilon=10^{-3}$
and $A=-1.32\times 10^{-4}$ the phase space presents two pairs of fixed
points (see Eq. (\ref{fixedpoints})). As we observe in Figs.
\ref{eigenvaluesphi1} and \ref{eigenvaluesphi2}, for such a combination of 
parameters, two of these fixed points are saddle and two are repelling
focus.

\begin{figure}[htb]
\centerline{
\includegraphics[angle=0, width=8.5cm, height=7cm]{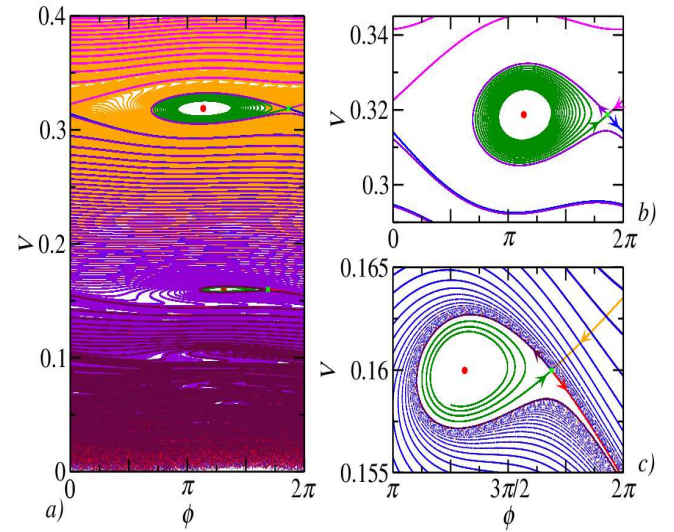}
}
\caption{
(Color online)
a) The figure illustrates the stable and unstable manifolds of the saddle points 
for $\varepsilon=10^{-3}$, $A=-1.32\times 10^{-4}$, 
$B=6.68\times 10^{-4}$ and $C=-1.72\times 10^{-4}$. 
In b) we have the manifolds near the fixed points with $i=1$ and in c) 
we have the manifolds near the fixed points with $i=2$.}
\label{A<0Manifolds}
\end{figure}

Figure \ref{A<0Manifolds}a) illustrates the manifolds of both saddle points.
We used $\varepsilon=10^{-3}$, $A=-1.32\times 10^{-4}$, 
$B=6.68\times 10^{-4}$ and $C=-1.72\times 10^{-4}$. 
Let us concentrate initially in the
manifolds associated to the saddle point with $i=1$.  Figure
\ref{A<0Manifolds}b) displays magnifications of these manifolds near the
corresponding saddle point. One of the stable manifolds comes from the high
energy portion of phase space and converges to the saddle point. The other
stable manifold comes from the proximity of the repelling fixed point with
$i=1$ and converges to the saddle point. The unstable manifolds produce
trajectories that evolve to the low energy region of the phase space. 
For these manifolds, the particle reaches the rest after some collisions. 

The behaviour discussed above for the manifolds of saddle fixed point with
$i=1$ applies also for the manifolds of the saddle point with $i=2$. An
amplification of these manifolds near the saddle point with $i=2$ is
presented in fig. \ref{A<0Manifolds}c). 

\begin{figure}[htb]
\centerline{
\includegraphics[angle=0, width=8.5cm, height=6cm]{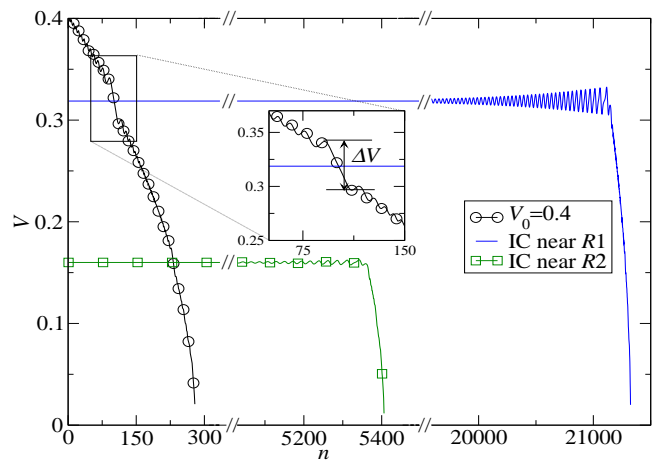}
}
\caption{
(Color online)
The plot illustrates the velocity decay of trajectories corresponding to three specific ICs. 
We used $\varepsilon=10^{-3}$, $A=-1.32\times 10^{-4}$, 
$B=6.68\times 10^{-4}$ and $C=-1.72\times 10^{-4}$. 
}
\label{NewFig.01}
\end{figure}

For $A<0$ all trajectories evolve to the low energy portion of phase space
leading the particle to reach the rest in the region between the walls.
Figure \ref{NewFig.01} illustrates this behavior for the trajectories of three
ICs. Two ICs were chosen near the repelling fixed points, named $R1$ and $R2$,
associated to $i=1$ and $i=2$, respectively. Each trajectory turns around the
corresponding repelling fixed point while the velocity oscillation increases. 
After a number of collisions the trajectories reach the region below the
saddle point and evolve to values of velocity that lead the particle to reach
the rest before a new collision. The third IC is $(\phi_0,V_0)=(0,0.4)$,
located above the fixed points. Above $R1$ the trajectory evolves decreasing
the value of $\phi$; below $R1$ the trajectory evolves increasing the value of
phase. The inset in Fig. \ref{NewFig.01} illustrates the variation of
velocity, $\Delta V$, during the incursion of the trajectory around $R1$. A
similar behaviour occurs when the trajectory passes around $R2$. After some
collisions, the value of velocity reaches the minimum value and the time
interval to the next collision diverges. Different ICs generate trajectories
with the same qualitative behaviours described above. For $A<0$ we have a
situation where the balance between the contribution of the electric field,
dissipation and amplitude of oscillation leads the particle to loose all its
kinetic energy. The exceptions to this rule are, obviously, the trajectories
in the stable manifolds and the fixed points. 

\begin{figure}[htb]
\centerline{
\includegraphics[angle=0, width=8.5cm, height=7cm]{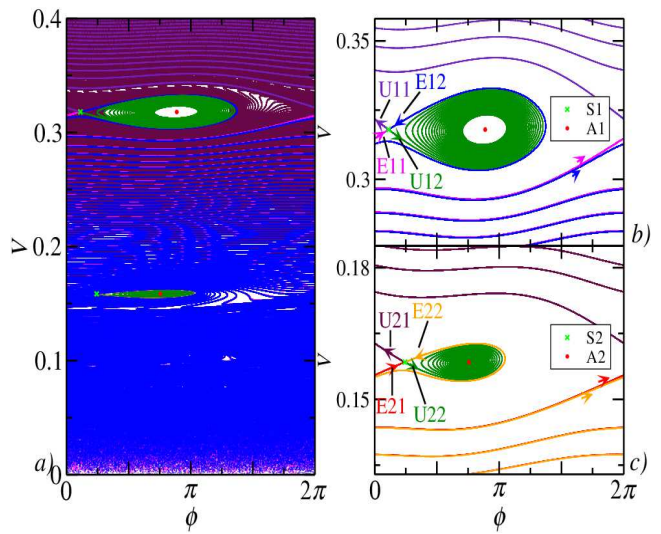}
}
\caption{ (Color online)
Figure a) displays the stable and unstable manifolds of saddle points for $\varepsilon=10^{-3}$,
$A=1.1\times 10^{-4}$, 
$B=1.9\times 10^{-4}$ and $C=1.09\times 10^{-4}$. 
Figures b) and c) illustrate these manifolds near the fixed points with
$i=1$ and the fixed points with $i=2$, respectively.}
\label{A>0Manifolds}
\end{figure}

For $A>0$, however, the manifolds organize the trajectories in phase space in a different way. 
For $\varepsilon=10^{-3}$ and $A=1.1\times 10^{-4}$ Eq. (\ref{fixedpoints}) 
furnishes two pairs of fixed points. Let us call the saddle and the 
attractor points for $i=1$ as $S1$ and $A1$, respectively. Similarly $S2$ and 
$A2$ are the saddle and attractor points for $i=2$. Figure
\ref{A>0Manifolds}a) illustrates the stable and unstable manifolds of the
saddle points. We used $\varepsilon=10^{-3}$, $A=1.1\times 10^{-4}$, 
$B=1.9\times 10^{-4}$ and $C=1.09\times 10^{-4}$. 
To simplify the
notation, let us call $U11$ and $U12$ the unstable manifolds of saddle point
$S1$, and $E11$, $E12$ the stable manifolds of $S1$. Similarly, $U21$ and
$U22$ are the unstable manifolds of $S2$, and $E21$, $E22$ are the stable
manifolds of $S2$. Figures \ref{A>0Manifolds}b) and c) illustrate these
manifolds near $S1$, $A1$ and $S2$, $A2$, respectively. 

The manifold $U11$ evolves to the high energy portion of phase space and
leads the particle to experience FA. The manifolds $U12$ and $U22$ evolve
towards the attractors $A1$ and $A2$, respectively. The stable manifolds of
both saddle points come from the low energy region of phase space. We
observe that $E11$ and $E12$ are close to each other below the saddle point
$S1$. Similarly, $E21$ is near $E22$ below $S2$. We observe that the region
of phase space located above $E12$ and below $E11$ forms a thin channel where
trajectories evolve until they pass below/near $S1$ and, after that, converge
asymptotically to the attractor $A1$. 

The region of phase space located above $E11$ and below $E12$ forms a large
channel where trajectories pass around $A1$ outside $E12$. These
trajectories access the region of high energy of the phase space and present
unlimited energy gain (FA). 

\begin{figure}[htb]
\centerline{
\includegraphics[angle=0, width=4.5cm, height=6.5cm]{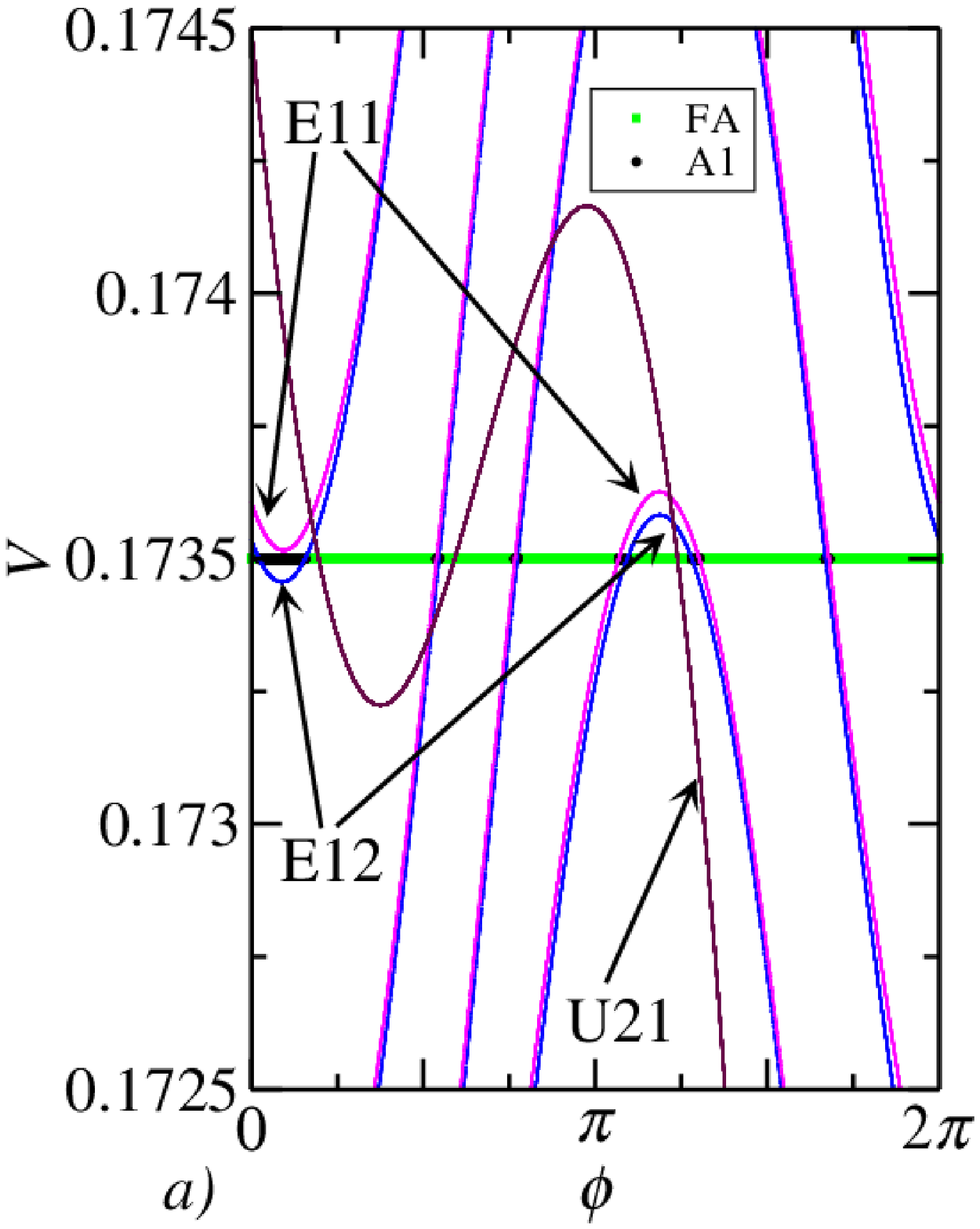}
\includegraphics[angle=0, width=4.5cm, height=6.5cm]{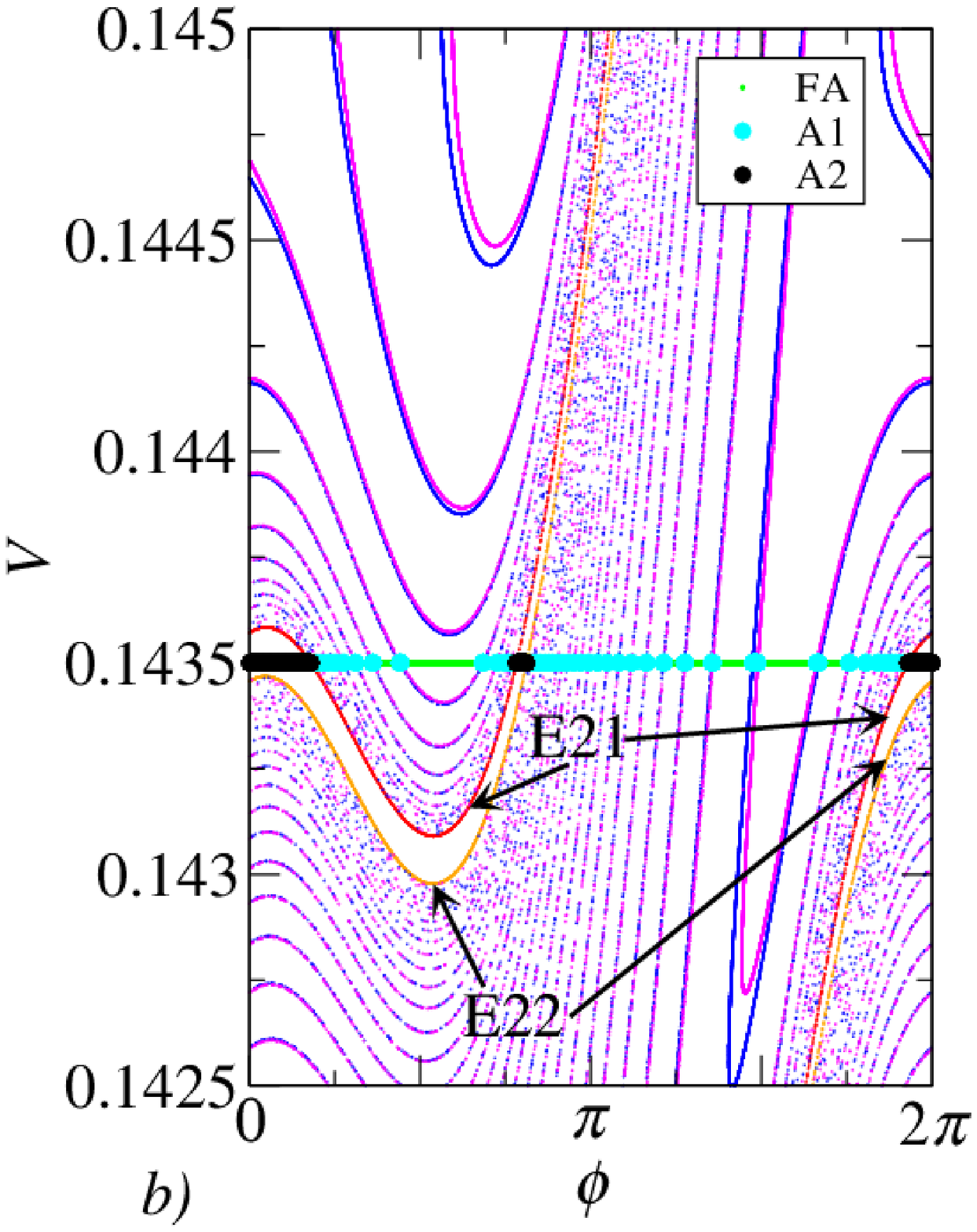}
}
\caption{(Color online) 
This figure shows the 
manifolds $E11$, $E12$ and $U21$ in a region a) between the fixed
points with $i=1$ and $i=2$, and b) in a region below the fixed points with
$i=2$. The parameters are $\varepsilon=10^{-3}$, $A=1.1\times 10^{-4}$, 
$B=1.9\times 10^{-4}$ and $C=1.09\times 10^{-4}$. 
}
\label{A1A2FA}
\end{figure}

To illustrate the behaviour discussed above, we define a set of $10^4$ ICs
with $V_0=0.1735$, $\phi_0$ randomly chosen in interval $(0,2\pi]$ and evolve
them in time. Figure \ref{A1A2FA}a) illustrates these ICs and the manifolds
$E11$, $E12$ and $U21$. The black points at $V_0=0.1735$ correspond to ICs
located in the thin channel above $E12$ and below $E11$ and they evolve to
$A1$. The gray (green) points at $V_0=0.1735$ are ICs in the large channel
above $E11$ and below $E12$ and present FA. We chose this value of initial
velocity because it is located above the fixed points with $i=2$ and below the
fixed points with $i=1$, a region where we observe heteroclinic intersections
of the unstable manifold $U21$ with the stable manifolds $E11$ and $E12$. 
Some points of $U21$ converge to the attractor $A1$ while most of them present
FA, depending on their locations in phase space with relation to $E11$ and
$E12$. 

\begin{figure}[htb]
\centerline{
\includegraphics[angle=0, width=8.5cm, height=7cm]{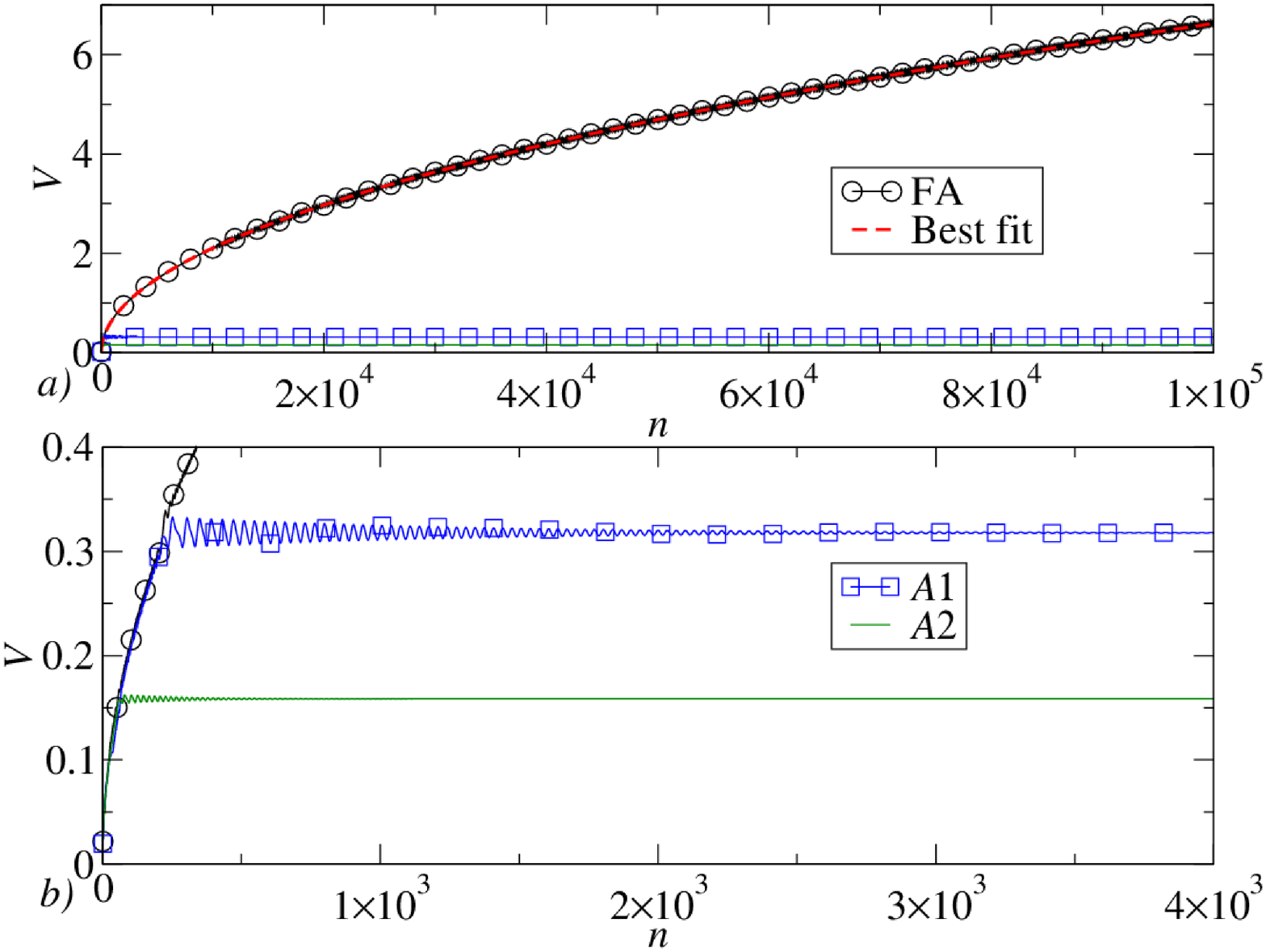}
}
\caption{
(Color online)
These plots illustrate the three asymptotic behaviors for 
$\varepsilon=10^{-3}$, $A=1.1\times 10^{-4}$, 
$B=1.9\times 10^{-4}$ and $C=1.09\times 10^{-4}$. 
Fig. b) is an amplification of Fig. a) in the region of small values of $n$ and $V$. 
}
\label{NewFig.02}
\end{figure}

Similar asymptotic behaviors are observed for the regions of phase space
limited by the stable manifolds $E21$ and $E22$. Figure \ref{A1A2FA}b)
illustrates such manifolds and the manifolds $E11$, $E12$ of Fig.
\ref{A1A2FA}a). We observe in Fig. \ref{A1A2FA}b) several incursions of $E11$
and $E12$ above and below the manifolds $E21$ and $E22$. As made before, we
defined a set of ICs with random values of $\phi_0$ with $V_0=0.1435$ and
evolved them. This value of $V_0$ was chosen because we are interested in the
behaviour of trajectories in the region below the fixed points with $i=2$,
where the stable manifolds $E21$ and $E22$ are located. The gray (cyan)
bullets correspond to ICs in the thin channel between $E12$ and $E11$. As
discussed before, the asymptotic behaviour of these ICs converges to $A1$. 
The black bullets denote ICs located in the thin channel above $E22$ and below
$E21$. All trajectories in this thin channel evolve to the attractor $A2$. The
small gray (green) points are in the large channel between $E11$ and $E12$
discussed above. Therefore, these ICs present FA. We must say that both the
thin and large channels formed by $E11$ and $E12$ assume a very stretched and
bended shape below the fixed points $A2$ and $S2$. Figure \ref{NewFig.02}
illustrates these asymptotic behaviours for three trajectories with
$V_0=10^{-3}$ and different values of $\phi_0$. Figure \ref{NewFig.02}a)
includes the best fit to the numerical data of the energy growth associated to
FA. The procedure furnishes that $V\propto n^{\gamma}$ with $\gamma\approx
1/2$ with good accuracy. Figure \ref{NewFig.02}b) is an amplification of the
portion corresponding to small values of $n$ and $V$ of Fig.
\ref{NewFig.02}a). In this figure we observe two trajectories evolving to the
spiral attractors $A1$ and $A2$. 

\begin{figure}[htb]
\centerline{
\includegraphics[angle=0, width=5cm, height=7cm]{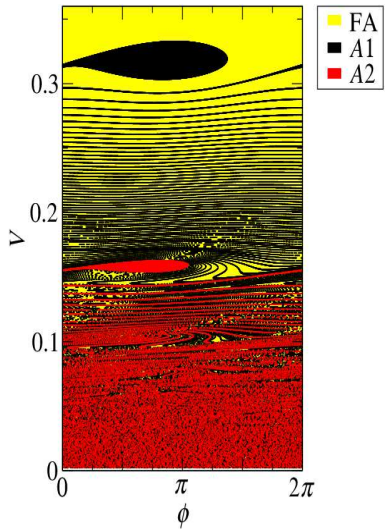}
}
\caption{(Color online) 
The black region in the plot corresponds to the basin of attraction of $A1$ and the grey (red)
points form the basin of attraction of spiral point $A2$. The light
grey (yellow) region corresponds to the ICs that presents FA. The parameters are
$\varepsilon=10^{-3}$, $A=1.1\times 10^{-4}$, $B=1.9\times 10^{-4}$ and $C=1.09\times 10^{-4}$. 
}
\label{BasinsAndFA}
\end{figure}

As the reasoning presented above for $V_0=0.1435$ and $V_0=0.1735$ can 
be extended for all the phase space, we defined a $10^3\times 10^3$ grid 
of ICs in phase space uniformly distributed in the intervals $0<\phi_0\le 2\pi$ 
and $0< V_0\le 0.36$ and let each initial condition to evolve in time
seeking for their final state. We present these ICs in Fig.
\ref{BasinsAndFA} where the different colors indicate the three possible
asymptotic behaviours of the trajectories. The black region corresponds to
ICs that converge to the attractor $A1$. Similarly, the gray (red) region
corresponds to the ICs that evolve to $A2$. The light gray (yellow) region
corresponds to ICs leading to FA. The previous discussion about the asymptotic
behaviour in terms of the manifolds is consistent with the shapes of the
basins of attraction of $A1$ and $A2$ and the channels associated to FA.

Let us present some technical information about the construction of the
manifolds. We used $10^4$ ICs in a maximum distance from the saddle points
of $10^{-3}$. The number of iterations changed depending on needs, sometimes
we used less than $300$ and sometimes $10^4$ iterations. 

The classification of the asymptotic behaviour of trajectories discussed 
about in Figs. \ref{A1A2FA} and \ref{BasinsAndFA} followed the procedure. 
We iterated the map for each IC until one of the
conditions was satisfied: (i) the trajectory converged to the attractor
$A1$, (ii) the trajectory converged to $A2$ or (iii) the value of velocity
reached the value $V=0.36$. We chose the value $V=0.36$ because it
guarantees the trajectory surrounded $A1$ at outside the manifold $E12$, 
had access to the high energy region of phase space and, therefore, 
it presents FA. 

The situations where all trajectories evolve to periodic 
and/or chaotic attractors located at some specific regions of phase space are usual.
The main point we are reporting here is the coexistence of attractors and
trajectories that present FA in the phase space of a dissipative system. In
what follow, we present the discussion on the complete model.

\subsection{The complete map}

The complete version takes into account the movement of the wall in the region
$[-\varepsilon,\varepsilon]$. The map of the complete version is described by
the two-dimensional map of the type
\begin{eqnarray}
T:\left\{
\begin{array}{l}
V_{n+1}=-V^{(n+1)}-2\varepsilon\sin\phi_{n+1}\\
\phi_{n+1}=(\phi_n+\Delta t_{n+1})\mod 2\pi
\end{array}
\right.,
\end{eqnarray}
where $V^{(n+1)}$ is the velocity of particle immediately before it collides
with the moving wall at instant $t_{n+1}=t_n+\Delta t_{n+1}$. The quantity
$t_{n+1}$ is the instant of collision $(n+1)$. The term $\Delta t_{n+1}$
is the smallest solution of equation $f(\Delta t_{n+1})=0$. The expressions of
$V^{(n+1)}$ and $f(\Delta t_{n+1})$ assume different forms depending on each
situation. Because there are several details to be regarded, we describe now
just some of them. 

Let us consider, for example, the situation where $A<0$ and $V_n>0$. The
quantity $X_s$ defines the position where the particle stops. If
$X_s=X_n-V_n^2/(2A)\le\varepsilon$, then we must determine if a collision
occurs (i) before or (ii) after the particle reaches the rest. In the case (i) we
have 
%case 7a begin
$V^{(n+1)}=V_n+A\Delta t_{n+1}$
and 
%case 7a
$f(\Delta t_{n+1})=X_n+V_n\Delta t_{n+1}+(A\Delta t_{n+1}^2)/2-\varepsilon\cos(\phi_n+\Delta t_{n+1})$, %case 7a
%case 7a end
while for the case (ii) we have 
%case 7b begin
$V^{(n+1)}=0$ 
and 
%case 7b
$f(\Delta t_{n+1})=X_s-\varepsilon\cos(\phi_n+\Delta t_{n+1})$. %case 7b
%case 7b end
If the quantity $X_s>\varepsilon$, then we must determine if a 
collision occurs before the particle leaves the region $[-\varepsilon,\varepsilon]$ using %case 8a
the equations of case (i) above. 
Note that these situations correspond to direct collisions, when the particle 
suffers successive collisions with the moving wall without leaving the
collision zone. 
If a collision does not occur and if $\varepsilon<X_s<1$, then the particle reaches the rest and %case 8b
its dynamics is over. 
If $X_s\ge 1$, then the particle hits the fixed wall and 
we have two possibilities: (I) the particle 
stops at some position $X\in(\varepsilon,1)$ and its dynamics dies or 
(II) the particle reaches the region $[-\varepsilon,\varepsilon]$. 
In the last case we must determine if 
(a) the particle stops before colliding the moving wall %10b
or (b) a collision occurs before the particle reaches the rest. %10a
For the case (a) we have 
%case 10b begin
$V^{(n+1)}=0$ 
and 
$f(\Delta t_{n+1})=X_s-\varepsilon\cos(p_{n+1})$,
where now $X_s=2-\varepsilon\cos\phi_n+V_n^2/(2A)$. 
%case 10b end
And for case (b) we have 
%case 10a begin
$V^{(n+1)}=V_{\varepsilon}^{-}-A(t_{n+1}-t_{\varepsilon}^{-})$ 
and 
$f(\Delta t_{n+1})=\varepsilon+V_{\varepsilon}^{-}(t_{n+1}-t_{\varepsilon}^{-})-(t_{n+1}-t_{\varepsilon}^{-})^2A/2-\varepsilon\cos(\phi_n+\Delta t_{n+1})$, 
where 
$
V_{\varepsilon}^{-}
=-\sqrt{V_n^2-2A(\varepsilon\cos\phi_n-2+\varepsilon)}
$
and
$
t_{\varepsilon}^{-}
=t_n-(V_n+V_{\varepsilon}^{-})/A.
$
%case 10a end
If $X_s<-\varepsilon$ we must use the equations of case (b). 
Note that the situations (a) and (b) correspond to indirect collisions, because 
the particle hits the fixed wall before it collides the moving wall. 

The other situations to be regarded in the complete version of the model
include $V_n<0$ and all the possible situations for $A>0$. We must also weigh
up the competition between the electric and the static/dynamic friction
forces. Depending on situation, the particle does not stop, but there are
situations where the particle stops only instantaneously and the signal of
velocity reverses. There are also situations where the particle remains in rest for a finite
time interval waiting for a collision with the moving wall and 
situations where the time interval between two collisions diverges, when the
particle stops between the walls forever. Our complete version includes also
locking phenomena, when both wall and particle move together until the instant
when the particle is launched. We do not explain in details all of the cases
here although the computational code takes into account all of the situations.

\begin{figure}[htb]
\centerline{
\includegraphics[angle=0, width=4.3cm, height=7cm]{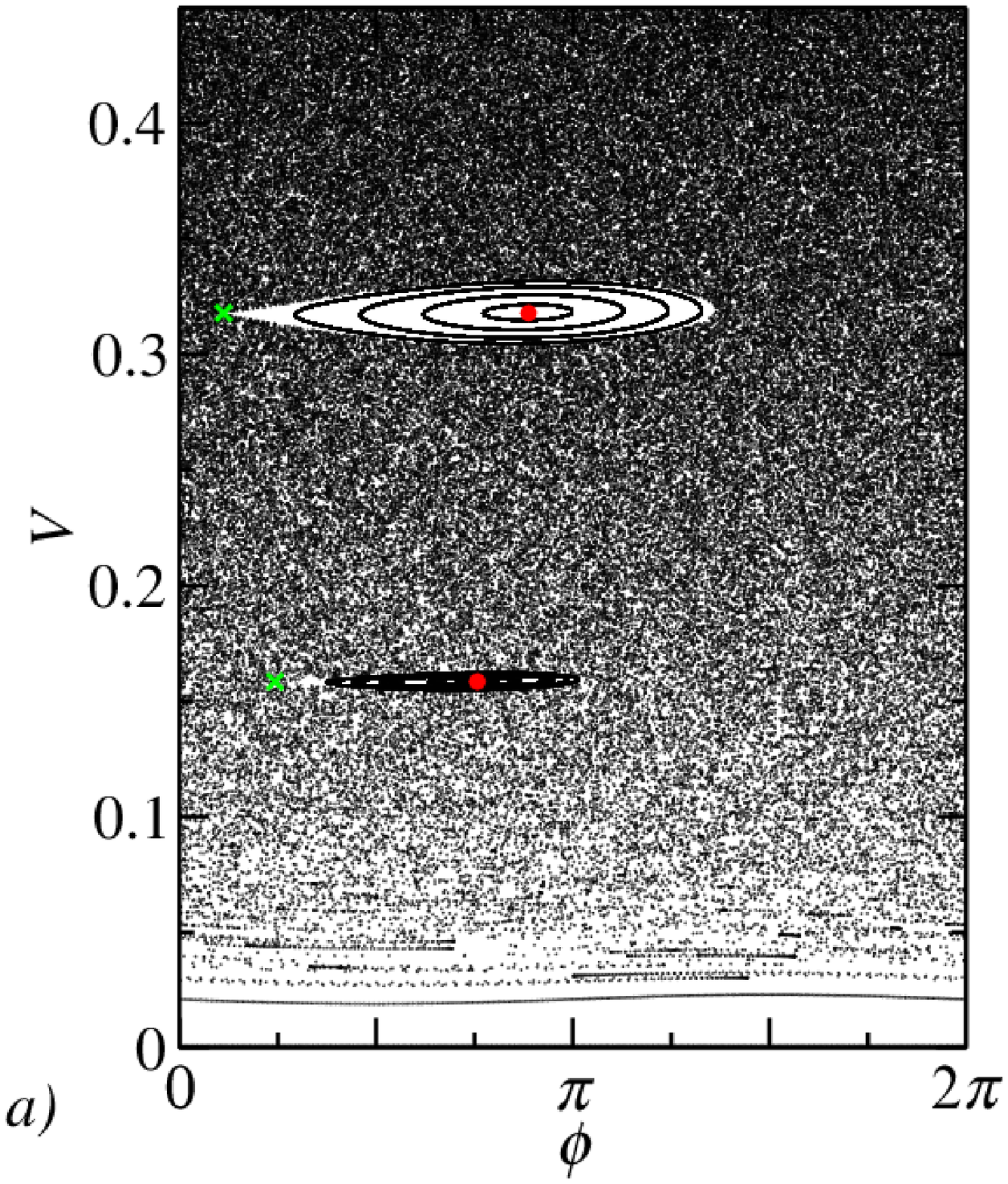}
\includegraphics[angle=0, width=4.3cm, height=7cm]{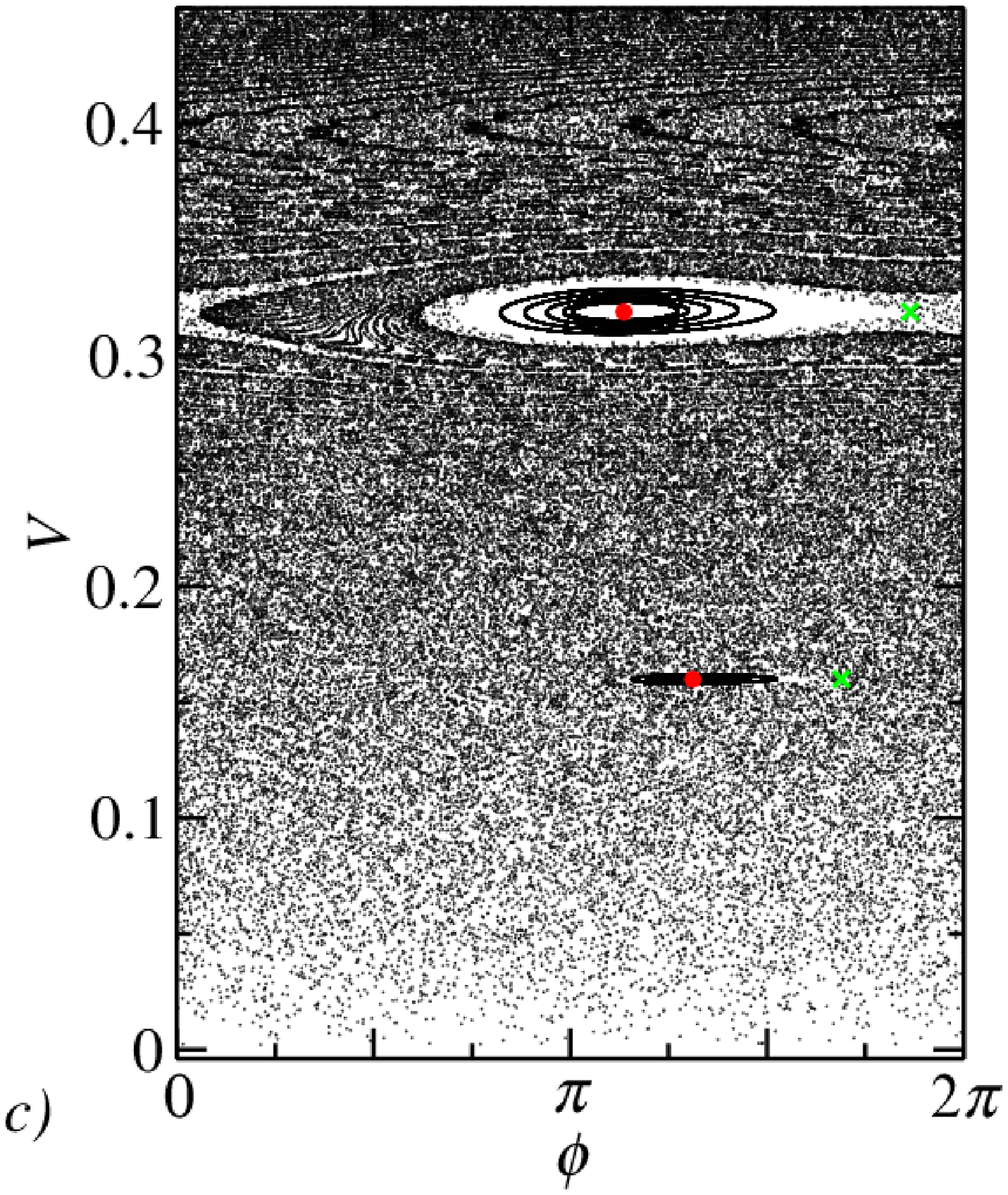}
}
\centerline{
\includegraphics[angle=0, width=4.3cm, height=3.5cm]{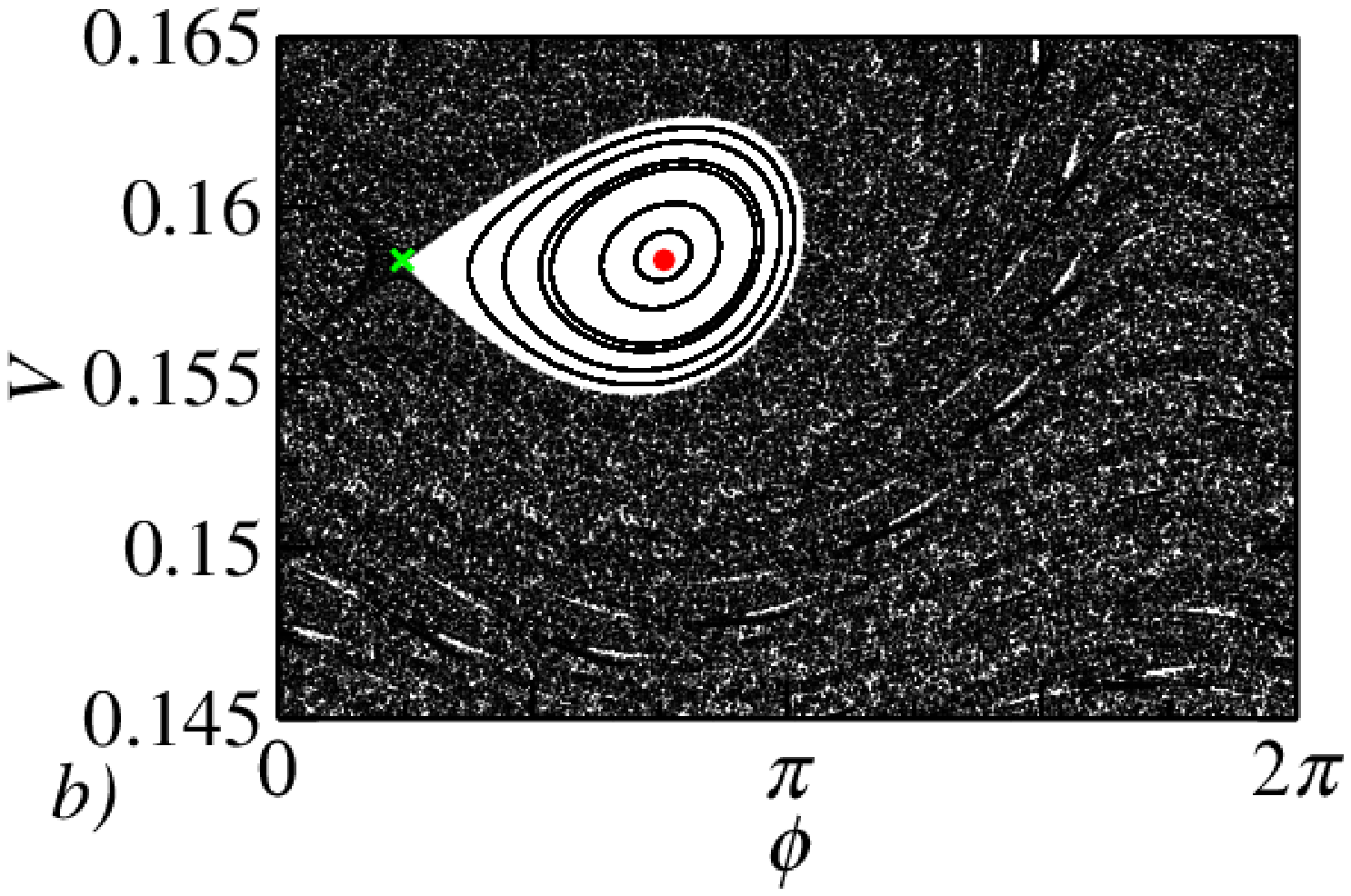}
\includegraphics[angle=0, width=4.3cm, height=3.5cm]{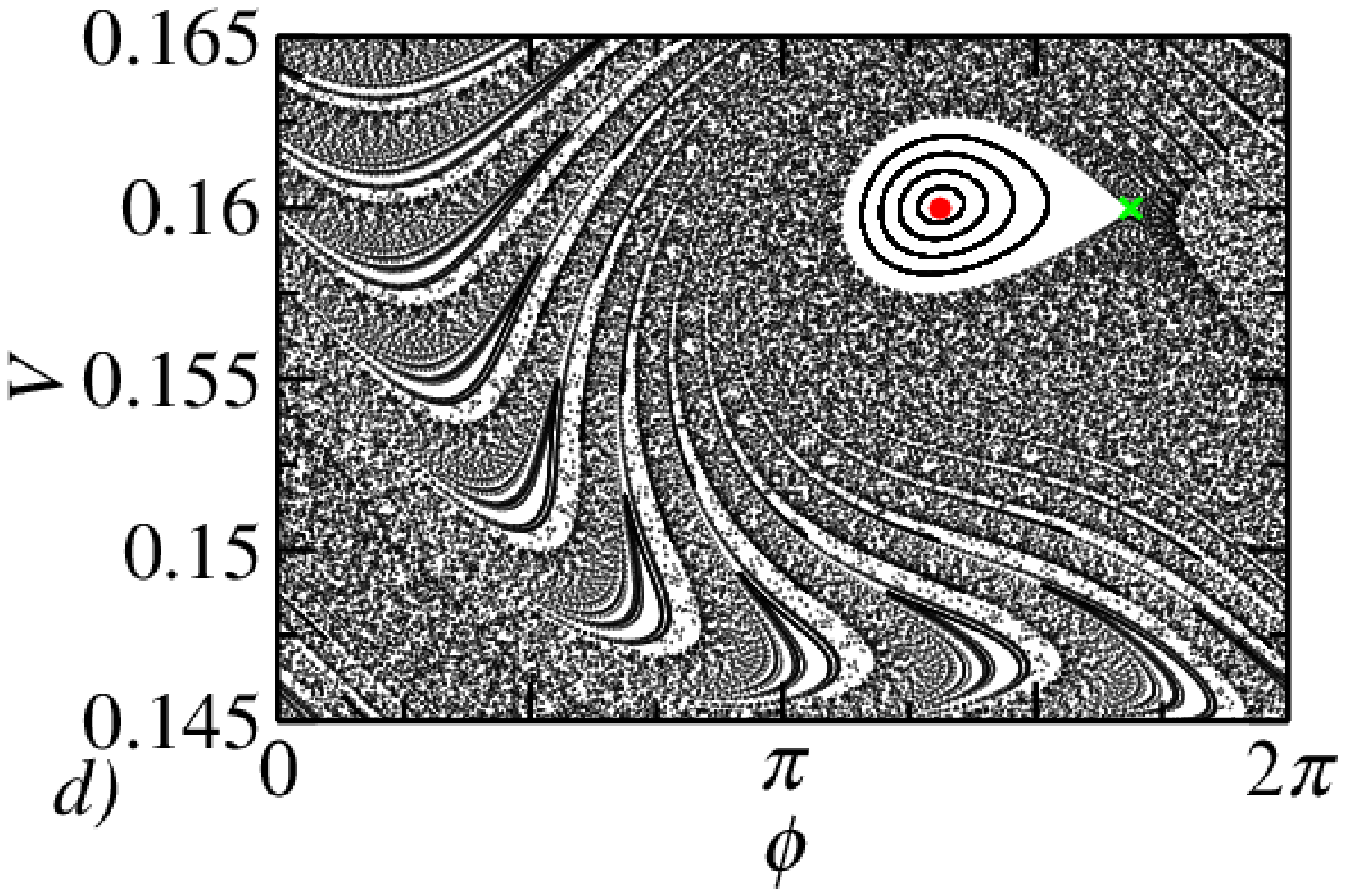}
}
\caption{
These plots illustrate the phase space of the complete model for two 
combinations of control parameters: 
a) $A=1.1\times 10^{-4}$, $B=1.9\times 10^{-4}$ and $C=1.09\times 10^{-4}$ and 
c) $A=-1.32\times 10^{-4}$, $B=6.68\times 10^{-4}$ and $C=-1.72\times 10^{-4}$. 
We used $\varepsilon=10^{-3}$ in both cases. 
The plots b) and d) correspond to magnifications of the regions of the small islands 
in plots a) and c), respectively. }
\label{A>0Complete}
\end{figure}

Figure \ref{A>0Complete} displays the phase space of complete model for both 
$A>0$ and $A<0$. Figure \ref{A>0Complete}a) illustrates the phase space of 
the complete model for $\varepsilon=10^{-3}$, $A=1.1\times 10^{-4}$, 
$B=1.9\times 10^{-4}$ and $C=1.09\times 10^{-4}$. 
These values of parameters 
are the same as those of Fig. \ref{A>0Manifolds}. We used a set of $5\times 10^3$ ICs
with $V_0=10^{-3}$ and different values of $\phi_0$ uniformly distributed in the 
interval $0<\phi_0\le 2\pi$. The trajectories of all these ICs
evolve to the high energy portion of phase space and present FA. We defined
two other sets of initial conditions, whose trajectories result in quasi-periodic 
orbits and generate the islands of regular motion observed 
in Fig. \ref{A>0Complete} a). The region near the island at $V\approx 0.16$ is
shown in Fig. \ref{A>0Complete}b). This result contrasts with the obtained for the simplified
model, which presents attracting focus for such a combination of parameters. 
Moreover, it is important to observe that the phase space presents a mixed
structure where regions of regular motion coexist with regions where
trajectories present FA. 

Figure \ref{A>0Complete}c) displays the phase space for
$\varepsilon=10^{-3}$, $A=-1.32\times 10^{-4}$, $B=6.68\times 10^{-4}$ and $C=-1.72\times 10^{-4}$, 
the same values used in Fig. \ref{A<0Manifolds}. 
We defined a set of $5\times 10^3$ ICs with $V_0=0.45$ and $0<\phi_0\le 2\pi$.
All these ICs evolve to the low energy region leading the trajectories
to reach the null velocity after a number of iterations. Therefore $V=0$
is an attractor for these trajectories. However, differently from the
simplified version, the phase space presents islands of regular motion for
$A<0$. Figure \ref{A>0Complete}d) illustrates the region near the smallest 
island located around $V=0.16$. For $A<0$ the complete version displays, therefore,
a coexistence of islands of regular motion with regions where trajectories
evolve to $V=0$. 
The $\times$ symbols and the circles denote the fixed points 
obtained from simplified map (\ref{SimplifiedMap}). 
As before, the former symbols correspond to saddle points. The last ones 
correspond to elliptic points in the complete model, instead the repelling or attracting 
nodes of the simplified map. 

The plot of the velocity as function of $n$ for $(\phi_0,V_0)=(0,10^{-3})$ 
and the parameters used in Figs. \ref{A>0Complete}a,b) is very similar to the curve that 
presents FA in Fig. \ref{NewFig.02}a) obtained for the simplified model. 
And the plot of a single trajectory with $V_0=0.4$ and the parameters of 
Figs. \ref{A>0Complete}c,d) is essentially the same observed in Fig. \ref{NewFig.01}. 
Therefore we do not include these plots of the complete model. 

Let us now discuss the above results. At first glance we may have a strange
feeling when we observe the coexistence of conservative and dissipative
behaviours in Figs. \ref{A>0Complete}c,d). However, the definition of a
`dissipative system' is not so clear \cite{book02}. Some people can say that
dissipation is associated to friction, which results in energy dissipation and
corresponds to non-modelled degrees of freedom. Other people can say that
dissipation corresponds to the situations where the volume of phase space is
not preserved or, in other words, the system is not described by the
Hamilton's equations. 

The recurrence theorem, which is a consequence of Liouville's theorem, 
states that, in a Hamiltonian system where the phase space is bounded, 
exists a finite neighbourhood of a point in phase space where trajectories 
emanates from and eventually return to this neighbourhood. 

In context of the results reported here, for the situations where the particle
reaches the rest between the walls, the time until the next collision
diverges. Therefore, in contrast to the hypothesis of the recurrence theorem,
the phase space of the system is not a bounded domain. The velocity axis is
also unbounded in vertical. The important point is that the theory protects
itself from an apparent paradox \cite{book02}; our results do not violate the
recurrence theorem. 

A similar coexistence of trajectories that present FA with KAM islands (Figs.
\ref{A>0Complete}a,b)) is observed for certain values of parameter in the
non-dissipative Fermi-Pustylnikov model \cite{BouncerOrig_Pustyl}. Regions of
regular dynamics were also reported in a simplified FUM with a drag force
proportional to the velocity of the particle \cite{diss.islands.01}.

As discussed before, the situations $A<0$ and $C<0$ can be interpreted as null
electric force ($qE=0$). In this case the parameters are given by $A=-\sigma
N/(\omega^2 l m)$, $B=\sigma N/(\omega^2 l m)$ and $C=-\sigma_s N/(\omega^2 l
m)$. Therefore, the discussed coexistence of islands of regular motion and an
attractor occurs in a FUM where a particle moves under friction in absence of
electric field. It is an interesting observation because it makes us to
believe that such a coexistence occurs in other systems as, for example, in
time dependent two-dimensional billiards. 

The main result we report here is the coexistence of an attractor located at 
$V=0$ with islands of regular motion. 

\subsection{Decay of energy: an analytical description for the case
$|A|/V^2\ll 1$}

As presented in previous sections, for $A<0$ both simplified and complete
models present trajectories that evolve to the low energy portion of the phase
space. We now present an analytical approximation to describe this velocity
decay. 

From the map of the simplified model, Eq. (\ref{SimplifiedMap}), we have for
$V\gg 2\varepsilon$ that $V_{n+1}\approx V_n\sqrt{1+\frac{4A}{V_n^2}}$. Given
an initial condition $V_0$ we have, after the first collision, $V_{1}\approx
V_0\sqrt{1+\frac{4A}{V_0^2}}$. If $4|A|/V_0^2\ll 1$ we evaluate a Taylor
expansion until fifth order in $A/V_0^2$ and obtain 
\begin{eqnarray}
\nonumber
V_1&\approx&V_0\left[1+2\frac{A}{V_0^2}-2\left(\frac{A}{V_0^2}\right)^2
+4\left(\frac{A}{V_0^2}\right)^3-10\left(\frac{A}{V_0^2}\right)^4
\right.
\\
\nonumber
&&
\left.
+28\left(\frac{A}{V_0^2}\right)^5+O\left(\left(\frac{A}{V_0^2}\right)^6\right)
\right],
\end{eqnarray}

When evaluating the expression of $V_2$ the quantity in the
square brackets in above equation appears as power with exponents
$-1,-3,-5,-7,-9$. Applying again the Taylor expansion until 5th order in
$A/V_0^2$ we obtain 
\begin{eqnarray}
\nonumber
V_2&\approx&V_0\left[1+4\frac{A}{V_0^2}-8\left(\frac{A}{V_0^2}\right)^2
+32\left(\frac{A}{V_0^2}\right)^3-160\left(\frac{A}{V_0^2}\right)^4
\right.
\\
\nonumber
&&
\left.
+896\left(\frac{A}{V_0^2}\right)^5+O\left(\left(\frac{A}{V_0^2}\right)^6\right)
\right]. 
\end{eqnarray}
Applying the same reasoning some times more, we obtain the following
approximation for $V_n$ 
\begin{eqnarray}
\label{TaylorApprox}
\nonumber
V_n\approx V_0\left[1+2n\frac{A}{V_0^2}-2n^2\left(\frac{A}{V_0^2}\right)^2
+4n^3\left(\frac{A}{V_0^2}\right)^3
\right.
\\
\left.
-10n^4\left(\frac{A}{V_0^2}\right)^4
+28n^5\left(\frac{A}{V_0^2}\right)^5+O\left(\left(\frac{A}{V_0^2}\right)^6\right)
\right]. 
\end{eqnarray}

\begin{figure}[t]
\centerline{
\includegraphics[angle=0, width=6cm, height=5cm]{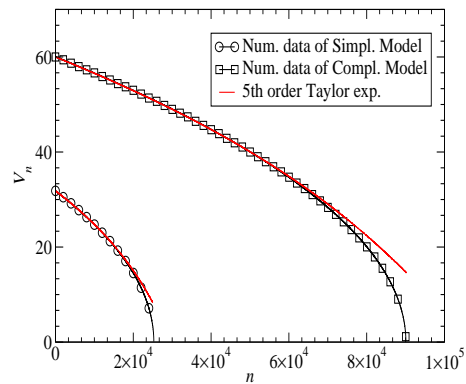}
}
\caption{
This plot illustrates the numerical and the analytical approximation of decay 
of velocity observed for $A<0$.}
\label{NumAnalyDecay}
\end{figure}

Figure \ref{NumAnalyDecay} displays the numerical data obtained by iterating
an IC for both simplified and complete versions. Figure \ref{NumAnalyDecay}
includes also the plot of the values of $V$ obtained from the above
approximation. For the simplified version we used $V_0=31.8$ and for the 
complete version we used $V_0=60$. The parameters used were
$\varepsilon=10^{-3}$, 
$A=-10^{-2}$, $B=1.2\times 10^{-2}$ and $C=-1.11\times 10^{-2}$ for both models. 
For small values of $n$ we observe a good agreement between
numerical data and the approximation given by Eq. (\ref{TaylorApprox}) for
both simplified and complete versions. The greater is the number of terms
in Taylor expansion the better is the agreement between numerical data and the
approximation for large values of $n$. 

Before the conclusions, let us discuss the results presented in the previous
sections. The energy of the particle is affected by three accelerating
mechanisms: i) the electric field, that most of time furnishes energy to the 
particle, ii) the dissipation on the surface, that drains the energy of the 
particle, and iii) the oscillating wall, that furnishes or absorbs energy from 
the particle depending on the phase. The friction force acts in the trajectory
of the particle continuously. The same rule applies for the electric force. 
The motion of the wall affects the energy of the particle at discrete time
instants, when the collisions occur. 

There are regions in phase space where the energy lost or received during the
travel between the walls comes into a dynamical equilibrium with the amount of
energy provided at the collisions instants with the oscillating wall in a such
a way that islands of regular motion are formed in the complete model. In
other words, there is a compensation between losing and gaining energy. These
islands are observed also for null electric field, where the energy dissipated
by the friction is balanced by the contribution of the oscillating wall.  

The simplified model furnishes a good approximation to the location of the
fixed points, given by Eq. (\ref{fixedpoints}). The dependence on parameter
$A$ in $V^*$ and $\phi^*$ expressions corresponds to the correction due to the
contributions of the electric and friction forces. For $A=0$ the expression
(\ref{fixedpoints}) furnishes the location of the fixed points in FUM. 

Regarding the results of the simplified map (\ref{SimplifiedMap}), we observe
in Fig. \ref{eigenvaluesphi2} that elliptic islands are observed for $A=0$,
when $|\Lambda_1|=|\Lambda_2|=1$. Taking the limit $A\rightarrow 0$ in the map
of simplified model we recover the situation where the particle moves
inertially between elastic collisions with the boundaries and we obtain the
map of FUM, as it is expected. The simplified FUM retains the nonlinearity
and several characteristics of the complete FUM, such as the mixed structure
of phase space, where KAM islands coexist with chaotic portions and invariant
spanning curves prevent trajectories of acquiring unlimited energy growth. 
The simplified version we study here preserves the occurrence of trajectories 
that present FA, for $A>0$, or trajectories that evolve to $V=0$, for $A<0$. 
However, although we regard small values of $\varepsilon$, when compared to the 
distance between the walls, the simplified model does not preserve the islands 
of regular motion observed in phase space of the complete model. 
Disregarding the small displacement of the particle inside the 
collision zone, defined by the interval $X\in[-\varepsilon,\varepsilon]$, 
affects the qualitative behavior of the trajectories near the fixed points. 

The stability of the islands of the complete model, Fig \ref{A>0Complete}, was
confirmed numerically for initial conditions iterated up to $10^9$ iterations.
Because islands of regular motion are observed in this dissipative system even
for null electric field, it is quite possible to one to find such structures in
the phase space of two-dimensional time dependent billiards, when the particle
slips on a rough surface. 

\section{Conclusions}
\label{conclusions}

We studied a version of Fermi-Ulam model where a metallic particle 
interacts with charged walls and with the field generated between them. 
Moreover, we regarded that the particle moves under action of a friction 
force due to its slip on a rough surface. The parameter $\varepsilon$ defines
the strength of nonlinearity and the parameters $A$, $B$ and $C$ represent
combinations of the electric and friction forces. We studied the dynamics of
the particle regarding a simplified and the complete versions of the model,
which present phase spaces with different structures. 

The phase space of the simplified version presents spiral repelling fixed 
points and trajectories that evolve to $V=0$, for $A<0$. In the other hand,
the phase space presents a coexistence of trajectories that evolve
asymptotically to attractor fixed points and trajectories that present Fermi
acceleration for $A>0$. The asymptotic behaviour of the trajectories is
described in terms of the location of initial points with relation to 
the channels formed by the stable manifolds of the saddle points. 

The phase space of complete model presents KAM islands coexisting with 
trajectories that i) present FA, when $A>0$, or ii) evolve to $V=0$, 
when $A<0$. We discuss that the coexistence of conservative and dissipative
behaviours observed for $A<0$ does not violate the recurrence theorem, because
the phase space is not a bounded domain. Moreover, we discuss that such
behaviour occurs even for null electric force and, therefore, this result
gives a hint of observation of this coexistence in other dynamical systems
where the particle moves under action of the friction force, including the
class of time dependent two-dimensional billiards. However, numerical
confirmation is needed. 

Finally, we obtained an analytical approximation to the velocity by evaluating
a Taylor expansion until 5th order in $A/V_0^2$ regarding the map of the
simplified model. So, we described the velocity decay observed in both
simplified and complete versions for $A<0$. The approximation is good for not
very long values of time, where the condition $|A|/V^2\ll 1$ is satisfied. 

\section*{ACKNOWLEDGEMENTS}
DGL thanks to CNPq and FAPESP. EDL thanks to FAPESP, CNPq and FUNDUNESP,
Brazilian agencies. This research was supported by resources supplied by the
Center for Scientific Computing (NCC/GridUNESP) of the S\~ao Paulo State
University (UNESP).

\end{document}